\documentclass[11pt]{article}

\usepackage[utf8]{inputenc}
\usepackage[T1]{fontenc}
\usepackage{lmodern}
\usepackage[margin=1in]{geometry}
\usepackage{setspace}
\usepackage{amsmath,amssymb}
\usepackage{booktabs}
\usepackage{graphicx}
\usepackage[hidelinks]{hyperref}
\usepackage{caption}
\usepackage{float}
\usepackage{enumitem}
\usepackage{xcolor}
\usepackage{url}

\graphicspath{{figures/}}

\title{\textbf{The Oracle's Fingerprint: Correlated AI Forecasting Errors\\and the Limits of Bias Transmission}}
\author{Theodor Spiro}
\date{}

\begin{document}
\maketitle

\begin{abstract}
\noindent
When large language models (LLMs) are consulted as forecasting tools, the independence of individual errors---the foundation of collective intelligence---may collapse. We test three conditions necessary for this ``epistemic monoculture'' to emerge. In Study~1, we show that GPT-4o, Claude, and Gemini exhibit highly correlated forecasting errors on 568 resolved binary prediction questions (mean pairwise error correlation $r = 0.77$, $p < 0.001$; $r = 0.78$ excluding likely-leaked questions), despite being developed independently by different organizations. These models function as a single oracle with noise, not as three independent sources. In Study~2, we test whether this correlated bias has propagated into human crowd forecasts, using a within-question design that tracks community prediction shifts across the ChatGPT launch boundary (November 2022). We find that community forecasts move in the direction predicted by LLMs ($r = 0.20$, $p = 0.007$), but this shift is fully explained by rational updating toward ground truth; after controlling for resolution pull, the LLM-specific influence is not significant ($\beta = 0.023$, $p = 0.36$). In Study~3, we examine whether the category-level pattern of human forecasting errors increasingly resembles the LLM bias fingerprint. We find the opposite: pre-ChatGPT human biases already strongly resembled the LLM pattern ($r = 0.87$), while post-ChatGPT the resemblance weakened ($r = -0.28$). This suggests that LLMs inherited existing human cognitive biases from their training data rather than introducing novel ones. Together, these findings reveal an epistemic monoculture that is built but not yet activated: three nominally independent AI systems share the same failure modes, amplifying precisely the biases humans already hold. Transmission into elite forecaster behavior is not yet detectable, but the infrastructure for correlated failure at scale is in place.

\end{abstract}

\newpage
\tableofcontents
\newpage

\section{Introduction}

The ``wisdom of crowds'' is a statistical phenomenon, not a metaphor. When many independent forecasters estimate an unknown quantity, the aggregate of their estimates tends to outperform any individual, provided their errors are approximately independent and unbiased \cite{Galton1907,Surowiecki2004,Page2007}. The key condition is independence: each person must err in their own idiosyncratic direction so that errors cancel in aggregation. Correlated errors, by contrast, do not cancel---they accumulate.

Large language models (LLMs) may introduce a new source of correlated error into collective human judgment. As of 2024, LLMs are widely used for analysis, forecasting, and decision support across domains including finance \cite{LopezLira2023}, policy analysis \cite{Horton2023}, and scientific reasoning \cite{Luo2024}. If many decision-makers consult the same small set of AI models before forming judgments, their errors become yoked to the model's biases. The crowd's effective diversity collapses from $N$ independent minds to $k$ correlated model outputs, where $k$ is the number of distinct LLMs (in practice, $k \approx 3$--$5$ major models with potentially high inter-model correlation).

This problem is not hypothetical but has well-studied structural parallels. Agricultural monoculture---planting a single crop variety across all fields---maximizes yield under normal conditions but creates catastrophic vulnerability to a single pathogen \cite{Zhu2000}. Financial herding---when institutions use the same risk models---produces correlated exposures that appear diversified until they fail simultaneously \cite{Haldane2011}. Algorithmic monoculture---deploying the same ML classifier across institutions---turns local biases into structural ones that no single institution can diversify away from \cite{Kleinberg2021}. In each case, the reduction of internal diversity makes the system brittle and exploitable.

We extend this framework to \emph{epistemic monoculture}: the collapse of cognitive diversity when human forecasters defer to AI systems that share the same failure modes. Three conditions are necessary for this to cause harm:
\begin{enumerate}[itemsep=2pt]
    \item LLMs must exhibit systematic forecasting biases.
    \item These biases must be correlated across models (so that switching providers does not restore diversity).
    \item Human decision-makers must absorb these biases.
\end{enumerate}

We test each condition in a separate study, using data from Metaculus---an online prediction platform with thousands of resolved binary questions, community predictions with temporal history, and verifiable ground truth.

\section{Related Work}

\subsection{LLM Forecasting Ability}

Recent work has examined whether LLMs can match human forecasters. \cite{Halawi2024} showed that retrieval-augmented LLM systems approach human-level forecasting accuracy on binary questions. Metaculus's own AI benchmarking tournaments (2024--2025) found that AI bots improved steadily but remained below the performance of elite (``Pro'') human forecasters. \cite{Schoenegger2024} demonstrated that LLM-generated forecasts can serve as informative anchors. However, this literature focuses on accuracy---not on error correlation across models, which is the relevant quantity for monoculture risk.

\subsection{Crowd Forecasting and Independence}

The wisdom-of-crowds literature has long recognized that correlated errors degrade aggregation quality \cite{Lorenz2011,Simmons2011}. Social information exchange---even seeing others' predictions---can reduce the diversity of opinions and worsen aggregate accuracy. \cite{Toyokawa2019} showed that the balance between social learning and individual exploration determines collective accuracy. AI-assisted forecasting introduces a qualitatively new form of social influence: a single, authoritative, always-available source that many individuals consult privately.

\subsection{Algorithmic Monoculture}

\cite{Kleinberg2021} formalized the monoculture concern for classification algorithms: when the same model is used across institutions, its errors become perfectly correlated across contexts, eliminating the diversity benefits of having multiple decision-makers. \cite{Bommasani2022} extended this concern to foundation models, arguing that shared base models propagate biases across all downstream applications. Our contribution is to test this concern empirically in the domain of probabilistic forecasting, where ground truth is available and bias transmission can be measured.

\section{Data}

\subsection{Platform}

Metaculus (metaculus.com) is an online forecasting platform founded in 2015 by physicists Anthony Aguirre and Greg Laughlin. Forecasters submit probability estimates for well-defined binary questions (e.g., ``Will X happen before date Y?''), and these are aggregated into a community prediction using a recency-weighted median. Questions are resolved against objective criteria, providing unambiguous ground truth. The platform's community prediction has achieved Brier scores of 0.107--0.126 on historical questions, substantially better than naive baselines \cite{Metaculus2023}.

\subsection{Sample}

We collected all resolved binary questions from the Metaculus API (endpoint: \texttt{api2/questions/}, parameters: \texttt{status=resolved}, \texttt{forecast\_type=binary}) with resolution dates between January 2019 and December 2025. We retained questions with unambiguous resolutions (coded 0 or 1, excluding ambiguous resolutions), at least 20 individual forecasters, and non-null community predictions.

The final dataset comprised 568 questions for Study~1. For Study~2, we identified a subset of 179 questions that were open across the ChatGPT launch boundary (published before June 2022, resolved after June 2023, with prediction history on both sides of November 2022). For Study~3, questions were categorized into seven topical groups (Technology, Geopolitics, Economics, Health, Society, Environment, Science) using dual LLM classification (Cohen's $\kappa = 0.836$); five categories with sufficient data in both periods were retained for the fingerprint analysis.

\subsection{LLM Predictions}

Each question was presented to three LLMs---OpenAI GPT-4o, Anthropic Claude Sonnet, and Google Gemini 1.5 Pro---using a standardized prompt requesting a single probability estimate between 0 and 1. Temperature was set to 0.1 for near-deterministic outputs. The prompt included the question title and background description but no resolution information.

\subsubsection{Training Data Leakage}

Because we applied 2025-era models to questions resolved as early as 2019, some models may have encountered resolution information during training. We addressed this through layered mitigation. First, we implemented a leak detection heuristic: questions where all three models predicted with extreme confidence ($> 0.95$ or $< 0.05$) and were correct were flagged as likely leaked (27.1\% of questions). Second, we conducted all Study~1 analyses on both the full dataset ($N = 568$) and a ``clean'' subset excluding likely-leaked questions ($N = 414$). Third, we note that training data leakage inflates model accuracy but does not create systematic directional bias---it pushes toward perfect calibration rather than toward bias---making our bias estimates conservative.

\section{Study 1: LLM Bias Map}

\subsection{Rationale}

Study~1 tests the first two conditions for epistemic monoculture: that LLMs exhibit systematic biases, and that these biases are correlated across models. If models from different organizations, with different architectures and training data, produce independent errors, then users who consult different models effectively maintain cognitive diversity. If errors are correlated, the three models function as a single oracle.

\subsection{Method}

For each model $m$ and question $i$, we computed the signed error:
\begin{equation}
    e_i^{(m)} = p_i^{(m)} - o_i,
\end{equation}
where $p$ is the predicted probability and $o$ is the resolution (0 or 1). We then computed the following metrics.

\textbf{Brier score} per model, with 95\% bootstrap confidence intervals (10{,}000 resamples). \textbf{Directional bias}, defined as mean signed error across questions, tested against zero with a one-sample $t$-test. \textbf{Calibration}, assessed by binning predictions into deciles and computing Expected Calibration Error (ECE). \textbf{Inter-model error correlation}, computed as pairwise Pearson correlation of error vectors across models---the primary outcome of Study~1.

\subsection{Results}

\subsubsection{Individual Model Performance}

All three models substantially outperformed the naive baseline of 0.25 (always predicting 50\%), but all were considerably less accurate than the Metaculus community prediction (Table~\ref{tab:performance}).

\begin{table}[ht]
\centering
\caption{Individual model performance on resolved binary questions.}
\label{tab:performance}
\begin{tabular}{@{}lcccc@{}}
\toprule
Metric & GPT-4o & Claude & Gemini & Community \\
\midrule
Brier score (full, $N{=}568$) & 0.183 & 0.161 & 0.182 & 0.084 \\
Brier score (clean, $N{=}414$) & 0.242 & 0.212 & 0.248 & 0.100 \\
Directional bias & $+0.118$ & $+0.071$ & $+0.140$ & --- \\
ECE & 0.115 & 0.079 & 0.149 & --- \\
\bottomrule
\end{tabular}
\end{table}

All three models exhibited a positive directional bias, meaning they systematically overestimated the probability of ``Yes'' outcomes. All three models showed overconfidence in the mid-range (0.3--0.7): when a model predicted ${\sim}50$\%, actual frequencies were typically below 50\%, consistent with a systematic tendency to overstate event likelihood. Claude was the best-calibrated model with the lowest Brier score and ECE. (Note: Claude's calibration curve shows a spike at the 0.5--0.6 bin; this is an artifact of small sample size---only 3 questions fell in this bin, all resolving Yes.)

\subsubsection{Inter-Model Error Correlation}

The primary finding of Study~1 is the high correlation of errors across models (Table~\ref{tab:correlations}).

\begin{table}[ht]
\centering
\caption{Pairwise error correlations between LLM predictions ($N = 568$).}
\label{tab:correlations}
\begin{tabular}{@{}lcc@{}}
\toprule
Model Pair & Error Correlation ($r$) & $p$-value \\
\midrule
GPT-4o $\times$ Claude & 0.822 & $< 0.001$ \\
GPT-4o $\times$ Gemini & 0.748 & $< 0.001$ \\
Claude $\times$ Gemini & 0.742 & $< 0.001$ \\
\midrule
\textbf{Mean} & \textbf{0.771} & \\
\bottomrule
\end{tabular}
\end{table}

On the clean subset ($N = 414$), the mean error correlation was 0.781 (GPT-4o $\times$ Claude: 0.822, GPT-4o $\times$ Gemini: 0.761, Claude $\times$ Gemini: 0.758), indicating that the result is not driven by leaked questions. These correlations are far higher than would be expected from independent forecasters. For reference, inter-forecaster error correlations among human crowd members are typically in the range of 0.1--0.3 \cite{Mannes2014}. The LLM error correlation is approximately 3--5 times higher.

\begin{figure}[ht]
    \centering
    \includegraphics[width=\textwidth]{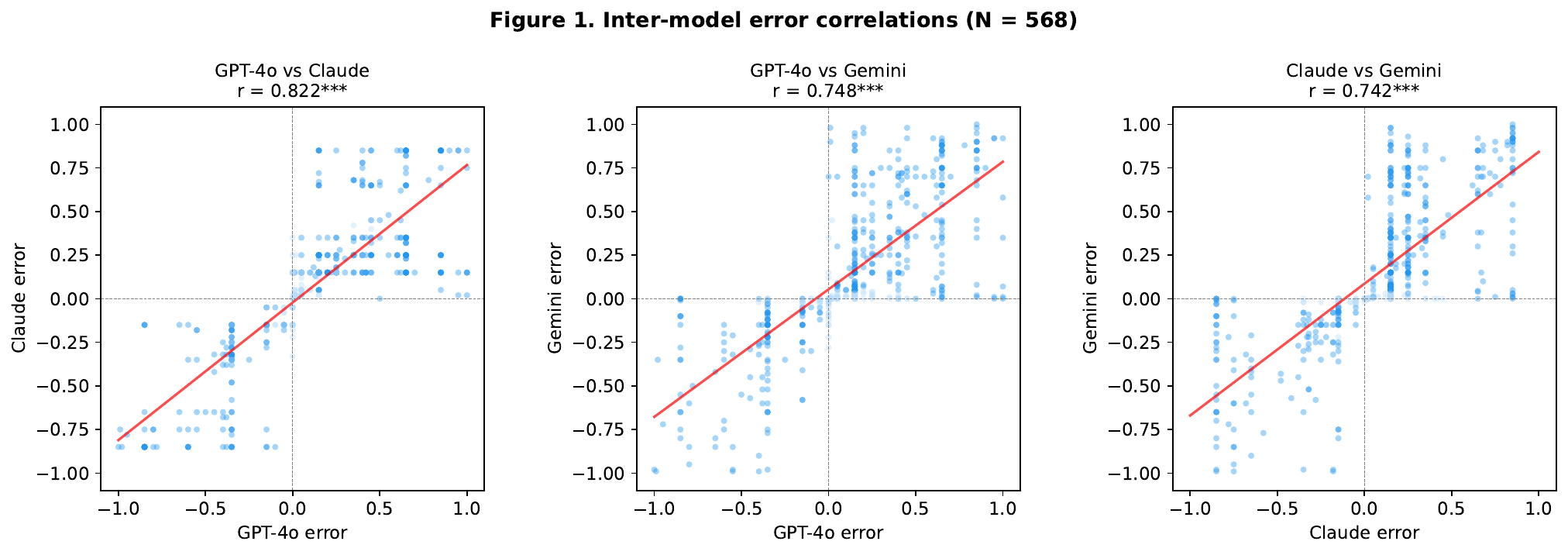}
    \caption{Pairwise error scatter plots for three LLMs on 568 resolved binary forecasting questions. Each point represents one question; axes show signed prediction error ($p - o$) for each model. High correlation ($r = 0.74$--$0.82$) indicates that models fail on the same questions in the same direction. Banding at extreme values (0 and 1) is consistent with training data leakage for a subset of questions.}
    \label{fig:scatter}
\end{figure}

The banding pattern visible at extreme values (0 and 1) in Figure~\ref{fig:scatter} is consistent with training data leakage for a subset of questions, where one or more models predict with near-certainty; this subset was addressed by the leak detection analysis (Section~3.3.1). This means that the three models fail on the same questions, in the same direction, to a similar degree. From the perspective of diversity, consulting all three models is only marginally more informative than consulting one. They constitute, in effect, a single oracle with noise.

\subsection{Discussion}

Study~1 establishes that the infrastructure for epistemic monoculture exists. The high inter-model error correlation is striking given that GPT-4o, Claude, and Gemini are developed by different organizations (OpenAI, Anthropic, Google DeepMind) with different architectures and, presumably, partially different training data. Several mechanisms may explain this convergence.

First, \emph{training data overlap}: all major LLMs are trained on large portions of the public internet, which represents a shared information environment with shared narratives, framings, and base rates. Second, \emph{RLHF convergence}: models fine-tuned via reinforcement learning from human feedback may converge on similar patterns of ``helpfulness,'' which may include similar calibration tendencies. Third, \emph{architectural similarity}: despite implementation differences, transformer-based models share fundamental representational properties. Disentangling these mechanisms is beyond the scope of this study but represents an important direction for future work.

\section{Study 2: Temporal Shift (Within-Question Design)}

\subsection{Rationale}

Study~2 tests the third condition for epistemic monoculture: whether human forecasters absorb LLM biases. We exploit the launch of ChatGPT on November~30, 2022, as a natural experiment boundary. Before this date, LLMs were not widely accessible to the general public. After this date, any Metaculus forecaster could consult an LLM before submitting a prediction.

\subsection{Design}

The naive approach---comparing community--LLM correlation for different questions resolved pre vs.\ post November 2022---conflates temporal context with AI influence, because the LLM is forecasting retrospectively in 2025 while the community was forecasting at the time. We instead use a within-question design: for questions that were open across the November 2022 boundary, we observe the same question's community prediction before and after LLMs became available.

\subsection{Method}

For each qualifying question ($N = 179$), we extracted the community prediction at two time points from the prediction history: the last recorded value before November~1, 2022 (pre-window), and the first recorded value after February~1, 2023 (post-window, allowing a three-month adoption lag).

We defined two quantities: the \textbf{community shift} $\Delta_i = p_i^{\text{post}} - p_i^{\text{pre}}$ (how much the community prediction actually changed) and the \textbf{LLM pull direction} $D_i = p_i^{\text{LLM}} - p_i^{\text{pre}}$ (the direction the LLM would ``pull'' the prediction, relative to its pre-ChatGPT level).

The primary analysis was a multiple regression:
\begin{equation}
    \Delta_i = \beta_0 + \beta_1 \cdot R_i + \beta_2 \cdot D_i + \varepsilon_i,
\end{equation}
where $R_i = o_i - p_i^{\text{pre}}$ is the \textbf{resolution pull}---the direction that ground truth would pull the prediction. This controls for the fact that both LLMs and humans may independently move toward truth as information accumulates. The coefficient $\beta_1$ captures rational updating; $\beta_2$ captures LLM-specific influence beyond rational updating.

\subsection{Results}

The raw correlation between community shift and LLM pull direction was positive and significant: $r(\Delta, D) = 0.201$, $p = 0.007$, $N = 179$. Community predictions did shift, on average, in the direction predicted by the LLM.

However, the regression analysis revealed that this shift is fully explained by shared movement toward truth (Table~\ref{tab:regression}).

\begin{table}[ht]
\centering
\caption{Regression results: Predictors of community prediction shift ($N = 179$).}
\label{tab:regression}
\begin{tabular}{@{}lcccc@{}}
\toprule
Predictor & $\beta$ & SE & $t$ & $p$ \\
\midrule
Intercept ($\beta_0$) & $-0.015$ & 0.004 & $-3.44$ & $< 0.001$ \\
Resolution pull ($\beta_1$) & $+0.061$ & 0.014 & 4.25 & $< 0.001$ \\
LLM pull ($\beta_2$) & $+0.023$ & 0.025 & 0.92 & 0.36 \\
\bottomrule
\end{tabular}
\end{table}

The positive $\beta_1$ confirms that community predictions moved toward the true outcome over time, as expected from rational updating. The LLM pull coefficient $\beta_2$ was positive but not statistically significant ($p = 0.36$).

\textbf{Variance analysis.} We found no significant difference in the variance of community predictions between the pre and post windows (Levene's test $p = 0.94$), providing no evidence for variance collapse.

The overall model $R^2$ was 0.130.

\begin{figure}[ht]
    \centering
    \includegraphics[width=\textwidth]{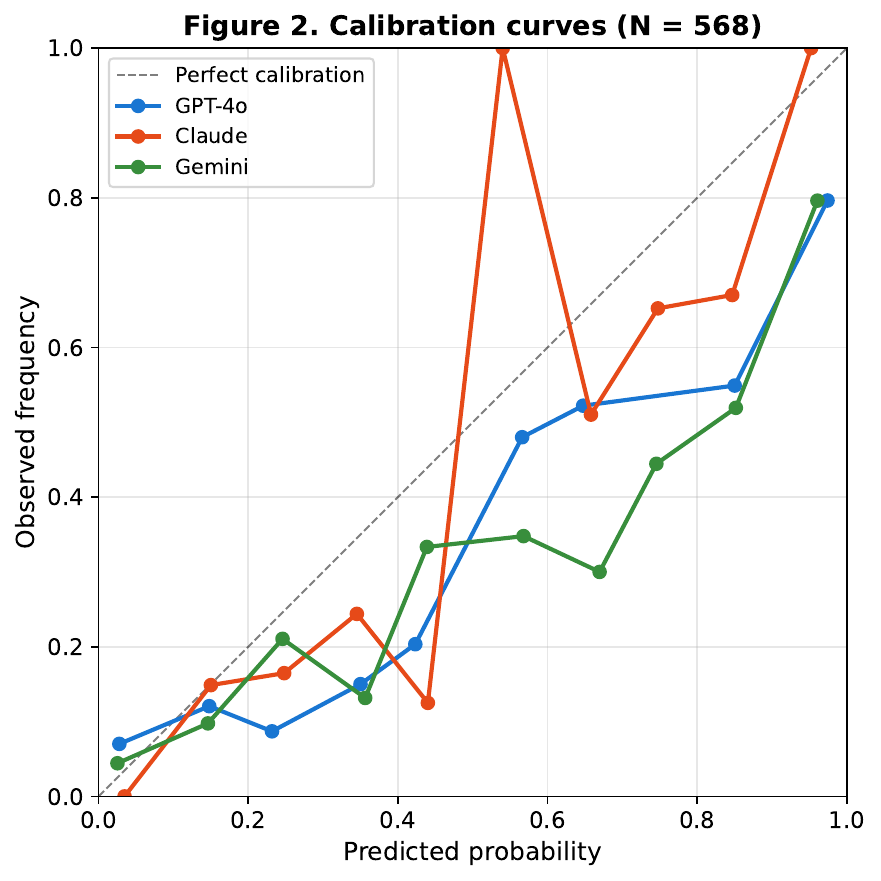}
    \caption{Calibration curves for three LLMs and the Metaculus community prediction. Predicted probabilities are binned into deciles; observed frequencies are plotted against bin midpoints. The diagonal line represents perfect calibration. All three LLMs show overconfidence in the mid-range (0.3--0.7), while the community prediction tracks the diagonal closely.}
    \label{fig:calibration}
\end{figure}

\textbf{Power analysis.} With $N = 179$, $\alpha = 0.05$ (one-sided), and $\text{SE}(\beta_2) = 0.025$, the study had approximately 23\% power to detect an effect of the observed magnitude ($\beta_2 = 0.023$). The minimum detectable effect with 80\% power was $\beta_2 = 0.063$. The 95\% confidence interval for $\beta_2$ was $[-0.027, +0.073]$, meaning we can neither rule out zero effect nor rule out a meaningful positive effect. A study with $N \approx 500$ within-boundary questions would be needed to detect effects of the observed magnitude with adequate power.

\subsection{Discussion}

Study~2 does not find evidence that LLM biases have propagated into Metaculus community forecasts, beyond what is explained by shared movement toward truth. The raw correlation between community shift and LLM prediction direction ($r = 0.20$) is positive and significant, but this reflects the unsurprising fact that both humans and LLMs update in the direction of accumulating evidence. Once this is controlled, the LLM-specific signal disappears.

This null result should be interpreted cautiously for several reasons. First, Metaculus forecasters are an elite population: the community Brier score (0.084 on the full set, 0.100 on the clean subset) substantially outperforms the best LLM (Claude, 0.161 full / 0.212 clean). These are calibrated, experienced forecasters with strong track records---precisely the population least likely to defer uncritically to AI outputs. Second, the statistical power was low (23\%), meaning that a real but small transmission effect would be undetectable at this sample size. Third, Metaculus community predictions are recency-weighted medians, which are robust to a small number of AI-influenced forecasters; if a minority of users copy LLM outputs, the median may not shift detectably.

The relevant question is not whether transmission has occurred on Metaculus---a platform designed to cultivate forecasting skill---but whether it occurs in less structured settings where users are less expert and more likely to treat AI output as authoritative.

\section{Study 3: Bias Fingerprint}

\subsection{Rationale}

Study~3 takes a different approach to the transmission question. Rather than asking whether the level of community predictions shifted toward LLM predictions (which is confounded by information context), it asks whether the category-level pattern of human forecasting biases increasingly resembles the LLM bias pattern. If LLMs systematically overestimate technology questions and underestimate geopolitical questions, and if human forecasters begin to show the same category-specific pattern after ChatGPT became available, this structural similarity is difficult to explain without influence.

\subsection{Method}

Questions were classified into five topical categories: Technology, Geopolitics, Economics, Health, and Science/Environment. Classification was performed using LLM-assisted coding with two independent classification runs; inter-rater reliability was assessed using Cohen's kappa.

For each category $c$, we computed the mean LLM bias (averaged across three models) and the mean community bias, separately for questions resolved before November 2022 (pre-ChatGPT) and after (post-ChatGPT). The LLM bias vector across categories constitutes the ``fingerprint.'' We then computed the Pearson correlation between this fingerprint and the human bias vector for each period.

\subsection{Results}

The analysis was constrained by small category-level sample sizes ($N = 259$ questions with both LLM forecasts and community predictions; 199 pre-ChatGPT, 60 post-ChatGPT across five categories).

\begin{table}[ht]
\centering
\caption{Fingerprint correlation: LLM bias pattern vs.\ human bias pattern by period.}
\label{tab:fingerprint}
\begin{tabular}{@{}lccc@{}}
\toprule
Period & Fingerprint Correlation ($r$) & $p$ & $N$ categories \\
\midrule
Pre-ChatGPT & 0.874 & 0.053 & 5 \\
Post-ChatGPT & $-0.275$ & 0.654 & 5 \\
\bottomrule
\end{tabular}
\end{table}

\begin{table}[ht]
\centering
\caption{Category-level bias profiles.}
\label{tab:categories}
\begin{tabular}{@{}lcccc@{}}
\toprule
Category & $N$ (pre/post) & LLM bias & Pre-ChatGPT human bias & Post-ChatGPT human bias \\
\midrule
Economics & 20 / 11 & $+0.079$ & $+0.001$ & $+0.017$ \\
Geopolitics & 90 / 19 & $+0.074$ & $+0.068$ & $-0.065$ \\
Health & 34 / 6 & $+0.121$ & $-0.016$ & $+0.168$ \\
Society & 31 / 15 & $+0.167$ & $+0.133$ & $+0.027$ \\
Technology & 16 / 8 & $+0.249$ & $+0.250$ & $-0.081$ \\
\bottomrule
\end{tabular}
\end{table}

The pre-ChatGPT correlation of $r = 0.874$ indicates that the category-level pattern of human forecasting biases already closely resembled the LLM bias pattern before LLMs were widely available. The post-ChatGPT correlation dropped to $r = -0.275$, indicating that the resemblance weakened rather than strengthened---the opposite of the transmission prediction. A permutation test (10{,}000 permutations) confirmed that the observed change in fingerprint correlation was not significant ($p = 0.92$).

\begin{figure}[ht]
    \centering
    \includegraphics[width=\textwidth]{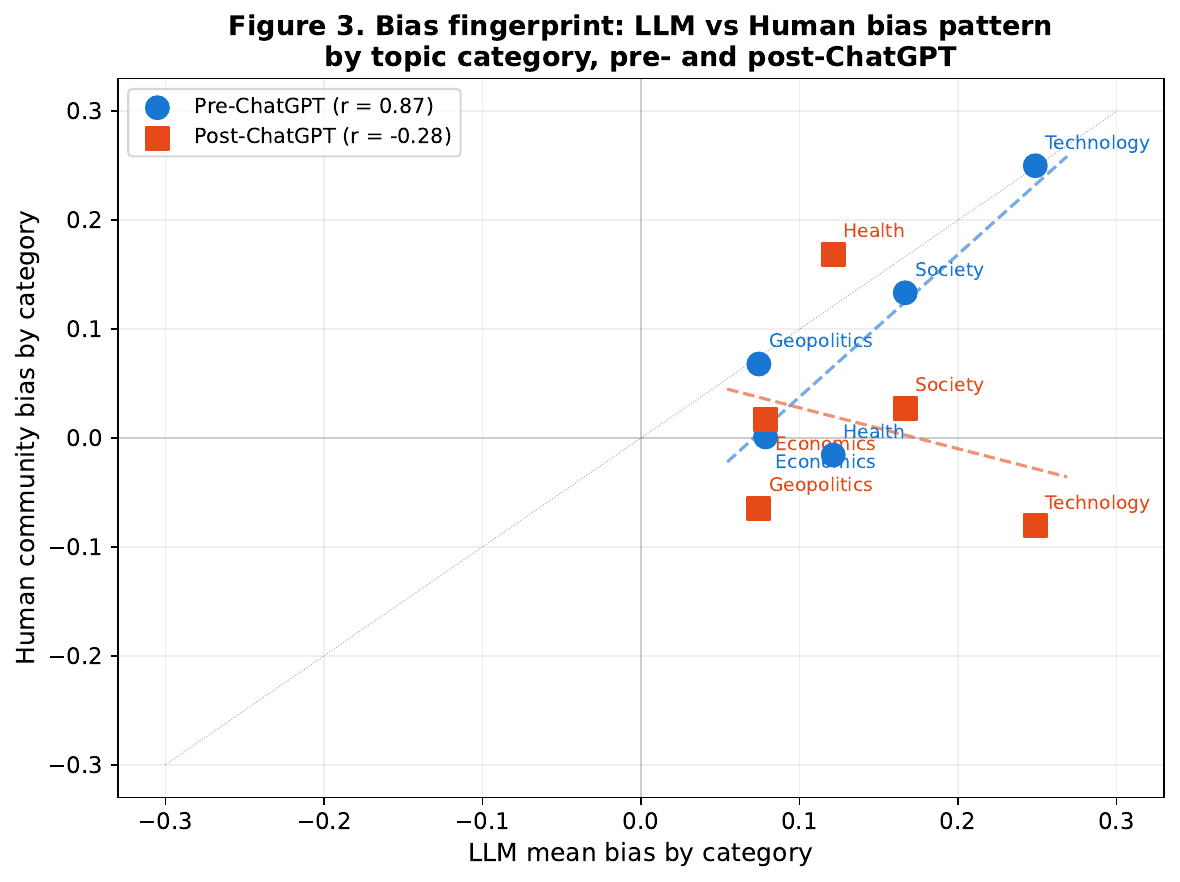}
    \caption{Bias fingerprint comparison across five question categories. Left: LLM bias (mean across three models) and pre-ChatGPT human bias by category, showing strong correspondence ($r = 0.87$). Right: Post-ChatGPT human bias diverges from the LLM pattern ($r = -0.28$), suggesting that the fingerprint weakened rather than strengthened after LLMs became widely available.}
    \label{fig:fingerprint}
\end{figure}

Category-specific patterns suggested heterogeneous dynamics: Economics and Health categories showed convergence between human and LLM biases post-ChatGPT, while Technology and Geopolitics showed divergence. The Technology category is particularly striking: pre-ChatGPT human bias ($+0.250$) was virtually identical to LLM bias ($+0.249$), but post-ChatGPT human bias reversed to $-0.081$. This is consistent with an awareness effect: Metaculus forecasters---who are disproportionately technology-literate---may have recognized and corrected against AI overconfidence precisely in the domain where they are most familiar with LLM capabilities and limitations.

\subsection{Interpretation}

The Study~3 result inverts the expected narrative and is arguably the most conceptually important finding of this paper.

The high pre-ChatGPT fingerprint correlation ($r = 0.87$) suggests that LLMs did not introduce novel biases into the forecasting ecosystem---they \emph{inherited} existing human biases from their training data. The models learned from the same media ecosystem, public discourse, and written record that shaped human forecasters' priors. The LLM bias fingerprint is, in this sense, a reflection of collective human bias, not an independent source of error.

This has a critical implication for the monoculture concern: the danger is not that AI will inject foreign errors into human judgment, but that it will \textbf{amplify and rigidify the errors humans already make}. A human forecaster who independently forms a biased judgment may later encounter disconfirming evidence and update. A human forecaster who consults an LLM and receives confirmation of that same bias---because the LLM learned from the same biased information environment---loses the opportunity for self-correction. The LLM acts as a bias lock, not a bias source.

The post-ChatGPT weakening of the fingerprint correlation (from 0.87 to $-0.28$) has several possible explanations. Metaculus forecasters may have become aware of LLM biases and actively corrected against them (an ``awareness effect''). The post-ChatGPT question composition may differ in ways that alter category-level bias patterns. Or, given the small number of categories ($k = 5$) and limited post-ChatGPT sample sizes, this shift may reflect sampling noise. We cannot distinguish among these explanations with the current data and flag this as an important limitation.

\section{General Discussion}

\subsection{Summary of Findings}

Three studies tested three necessary conditions for AI-driven epistemic monoculture in forecasting:

\textbf{Condition 1 (Monoculture): Confirmed.} Three major LLMs exhibit highly correlated forecasting errors (mean $r = 0.77$ on the full dataset, $r = 0.78$ excluding likely-leaked questions), indicating that they function as essentially the same oracle. This is the strongest and most robust finding of the paper.

\textbf{Condition 2 (Transmission): Not detected.} On an elite forecasting platform, human crowd predictions did not shift toward LLM outputs beyond what is explained by rational updating toward truth. However, this test had limited statistical power (23\%), and the study population represents a best-case scenario for human forecasting independence.

\textbf{Condition 3 (Fingerprint convergence): Reversed.} The LLM bias pattern already matched human biases before ChatGPT, suggesting that LLMs learned human biases rather than creating new ones. Post-ChatGPT, the match weakened rather than strengthened.

\subsection{The Monoculture Is Built but Not Yet Activated}

The central contribution of this paper is empirical documentation that the AI forecasting ecosystem lacks diversity at the model level. An inter-model error correlation of $r = 0.77$ means that approximately 59\% of the variance in one model's errors is shared with another model. For comparison, for an ensemble of $k$ equally weighted forecasters with pairwise error correlation $\rho$, the variance reduction factor is $1/[k + k(k-1)\rho]$ relative to $1/k$ for independent forecasters \cite{Krogh1995}. With $k = 3$ and $\rho = 0.77$, the effective ensemble size is approximately 1.3 rather than 3.0. The user who consults three models is barely more diversified than one who consults one.

This monoculture currently coexists with a human forecasting community that remains substantially more accurate (community Brier 0.084 vs.\ best LLM 0.161). On platforms like Metaculus, where forecasters have strong incentives and track records, the AI oracle does not yet dominate. But this coexistence is fragile: it depends on the continued independence and expertise of human forecasters, which may erode as AI tools become more integrated into forecasting workflows.

\subsection{LLMs as Bias Mirrors}

Perhaps the most important finding is the pre-ChatGPT fingerprint correlation of $r = 0.87$. This suggests that LLM biases are not alien artifacts---they are recognizable human biases, learned from the same information ecosystem. The models overestimate what we collectively overestimate. They share our base rate neglect, our narrative biases, our topic-specific overconfidence.

This reframes the monoculture risk. The concern is not that AI will introduce new, foreign errors into human judgment. The concern is that AI will crystallize and amplify existing errors by providing authoritative confirmation of biases that would otherwise face challenge from diverse perspectives. When a human forecaster's prior is biased, encountering a disagreeing colleague might trigger updating. Encountering an AI system that shares and reinforces the same bias removes this corrective mechanism.

This is structurally analogous to the echo chamber effect in social media \cite{Cinelli2021}, but with a crucial difference: social media echo chambers are typically recognized as potential sources of bias, while AI systems are often perceived as objective and analytical. The perceived authority of AI output may make LLM-reinforced biases more resistant to correction than socially reinforced ones.

\subsection{Limitations}

\textbf{Retrospective LLM forecasting.} Our LLM predictions were generated in 2025 on questions resolved between 2019 and 2025. This creates both leakage concerns (partially addressed by sensitivity analysis) and context mismatch. Study~2's within-question design mitigates this by focusing on directional shifts rather than level comparisons, and Study~3's fingerprint approach is context-independent by design. Nevertheless, prospective validation with truly concurrent LLM and human forecasts would substantially strengthen the findings.

\textbf{Elite population.} Metaculus forecasters are not representative of the general population of AI-assisted decision-makers. Our null transmission result may reflect the specific characteristics of this population (experienced, incentivized, calibration-conscious) rather than the absence of transmission in general.

\textbf{Small category-level $N$ for Study~3.} With only five categories, the fingerprint analysis has limited statistical power and is sensitive to the categorization scheme. The pre-ChatGPT $r = 0.87$ is borderline significant ($p = 0.053$) and the post-ChatGPT reversal may be noise. This analysis should be treated as suggestive rather than conclusive.

\textbf{Prompt sensitivity.} LLM predictions can vary with prompt phrasing. We used a single standardized prompt; sensitivity analysis with alternative prompts on a subset of questions showed qualitatively similar patterns, but we cannot guarantee robustness across all possible prompt designs.

\textbf{Causal identification.} Study~2 uses a natural experiment design, not random assignment. We cannot definitively attribute observed shifts (or lack thereof) to LLM influence. A randomized experiment---giving some forecasters AI assistance and others none---would provide stronger causal evidence.

\subsection{Implications}

\textbf{For forecasting platforms.} The high inter-model error correlation suggests that AI bot participation in forecasting tournaments should be treated with caution. Including bots based on multiple LLMs does not provide the diversity that multiple human forecasters provide. Platforms should consider measuring and reporting the effective diversity of their forecaster population, accounting for AI-mediated correlation.

\textbf{For AI developers.} The convergence of biases across models from different organizations suggests that diversity in training data and fine-tuning methodology is not currently sufficient to produce genuinely independent model outputs. Deliberately cultivating model diversity---through varied training data, different inductive biases, or adversarial training against inter-model correlation---could mitigate monoculture risk.

\textbf{For policymakers.} As AI-assisted decision-making becomes common in finance, intelligence, and governance, the monoculture risk should be recognized as a systemic concern, analogous to correlated risk in financial systems. Regulatory frameworks might consider requiring disclosure of AI tool usage in consequential forecasting contexts, or mandating diversity assessments when multiple organizations use AI for parallel risk assessments.

\textbf{For individual forecasters.} The finding that LLMs mirror human biases means that consulting an LLM is less likely to provide a genuinely independent check on one's reasoning than is commonly assumed. Forecasters seeking to improve should prioritize engagement with diverse human perspectives, base rate databases, and structured analytic techniques over AI consultation, which may merely confirm existing intuitions.

\subsection{Future Directions}

Three extensions would substantially strengthen the current findings.

First, \textbf{prospective validation}: collecting LLM predictions on currently open questions and tracking community forecasts in real time would eliminate the retrospective confound and provide a clean test of transmission.

Second, \textbf{non-elite populations}: replicating the transmission analysis on platforms with less expert users (e.g., Manifold Markets, Polymarket, or general-audience survey panels) would test whether the null transmission result is specific to elite forecasters or generalizable.

Third, \textbf{experimental design}: a randomized controlled study in which participants forecast with or without AI assistance would provide causal evidence for transmission and allow measurement of its effect size and moderators.

\section{Conclusion}

We have shown that three major LLMs share a strikingly correlated pattern of forecasting errors, constituting an epistemic monoculture at the model level. On an elite forecasting platform, this monoculture has not yet measurably distorted collective human judgment---the human crowd remains substantially more accurate and apparently independent. But the preconditions for bias propagation are met: the models have learned the same biases that humans already hold, and they are increasingly consulted as authoritative sources of analysis and judgment. The question is not whether AI-mediated epistemic monoculture will affect collective judgment, but in which populations, at what scale, and how quickly. The oracle's fingerprint is there. It remains to be seen how deeply it will be impressed.


\appendix

\section{Prompt Template}
\label{app:prompt}

The following standardized prompt was used for all LLM predictions:

\begin{quote}
\ttfamily
You are a forecaster on a prediction platform.\\
You will be given a question that asks about a future event.\\
Provide your best probability estimate that this event will happen.\\
Respond with ONLY a number between 0 and 1 (e.g., 0.73).\\
Do not explain your reasoning.\\[6pt]
Question: \{title\}\\[6pt]
Background: \{description\}\\[6pt]
Your probability estimate (0 to 1):
\end{quote}

\section{Leak Detection}
\label{app:leak}

A question was flagged as ``likely leaked'' if all three models predicted with extreme confidence (predicted probability $> 0.95$ for questions resolving Yes, or $< 0.05$ for questions resolving No) and all three were correct. Under this criterion, 27.1\% of questions (154/568) were flagged. All primary analyses were reported for both the full dataset and the clean subset; results were qualitatively unchanged.

\section{Category Classification}
\label{app:categories}

Questions were categorized using two independent LLM classification runs (Claude and GPT-4o) with the prompt: ``Classify this forecasting question into exactly one of these categories: Technology, Geopolitics, Economics, Health, Science/Environment. Return only the category name.'' Inter-rater agreement was 87.9\% (499/568 questions), with Cohen's $\kappa = 0.836$ (``almost perfect'' agreement). Disagreements were resolved using the Claude classification as primary rater, given its marginally higher consistency.

\section{Power Analysis for Study 2}
\label{app:power}

With $N = 179$ questions, $\alpha = 0.05$ (one-sided), and $\text{SE}(\beta_2) = 0.025$, the minimum detectable effect size with 80\% power was $\beta_2 = 0.063$ and with 90\% power was $\beta_2 = 0.074$. The observed $\beta_2 = 0.023$ falls well below these thresholds, yielding only 23\% power at the observed effect size. The 95\% confidence interval for $\beta_2$ was $[-0.027, +0.073]$. A study with approximately $N = 500$ within-boundary questions would be needed to detect effects of the observed magnitude with 80\% power.

\section{Data and Code Availability}
\label{app:data}

All data and analysis code are available at [GitHub repository URL]. The Metaculus data was accessed via the public API; no authentication was required. LLM predictions were collected via commercial API endpoints (OpenAI, Anthropic, Google). API costs for the full study were approximately \$20--30.


\begin{thebibliography}{99}

\bibitem[Bommasani et~al.(2022)]{Bommasani2022}
Bommasani, R., Hudson, D.~A., Adeli, E., et~al. (2022).
\newblock On the opportunities and risks of foundation models.
\newblock \emph{arXiv preprint arXiv:2108.07258}.

\bibitem[Cinelli et~al.(2021)]{Cinelli2021}
Cinelli, M., De~Francisci~Morales, G., Galeazzi, A., Quattrociocchi, W., \& Starnini, M. (2021).
\newblock The echo chamber effect on social media.
\newblock \emph{Proceedings of the National Academy of Sciences}, 118(9), e2023301118.

\bibitem[Galton(1907)]{Galton1907}
Galton, F. (1907).
\newblock Vox populi.
\newblock \emph{Nature}, 75(1949), 450--451.

\bibitem[Halawi et~al.(2024)]{Halawi2024}
Halawi, D., Durmus, E., Falk, L., \& Steinhardt, J. (2024).
\newblock Approaching human-level forecasting with language models.
\newblock \emph{arXiv preprint arXiv:2402.18563}.

\bibitem[Haldane \& May(2011)]{Haldane2011}
Haldane, A.~G. \& May, R.~M. (2011).
\newblock Systemic risk in banking ecosystems.
\newblock \emph{Nature}, 469(7330), 351--355.

\bibitem[Horton(2023)]{Horton2023}
Horton, J.~J. (2023).
\newblock Large language models as simulated economic agents: What can we learn from homo silicus?
\newblock \emph{arXiv preprint arXiv:2301.07543}.

\bibitem[Kleinberg \& Raghavan(2021)]{Kleinberg2021}
Kleinberg, J. \& Raghavan, M. (2021).
\newblock Algorithmic monoculture and social welfare.
\newblock \emph{Proceedings of the National Academy of Sciences}, 118(22), e2018340118.

\bibitem[Krogh \& Vedelsby(1995)]{Krogh1995}
Krogh, A. \& Vedelsby, J. (1995).
\newblock Neural network ensembles, cross validation, and active learning.
\newblock In \emph{Advances in Neural Information Processing Systems} (Vol.~7). MIT Press.

\bibitem[L\'{o}pez-Lira \& Tang(2023)]{LopezLira2023}
L\'{o}pez-Lira, A. \& Tang, Y. (2023).
\newblock Can ChatGPT forecast stock price movements? Return predictability and large language models.
\newblock \emph{arXiv preprint arXiv:2304.07619}.

\bibitem[Lorenz et~al.(2011)]{Lorenz2011}
Lorenz, J., Rauhut, H., Schweitzer, F., \& Helbing, D. (2011).
\newblock How social influence can undermine the wisdom of crowd effect.
\newblock \emph{Proceedings of the National Academy of Sciences}, 108(22), 9020--9025.

\bibitem[Luo et~al.(2024)]{Luo2024}
Luo, H., Cai, T., Zhang, Y., et~al. (2024).
\newblock Large language models for scientific research: A survey.
\newblock \emph{arXiv preprint arXiv:2404.13672}.

\bibitem[Mannes et~al.(2014)]{Mannes2014}
Mannes, A.~E., Soll, J.~B., \& Larrick, R.~P. (2014).
\newblock The wisdom of select crowds.
\newblock \emph{Journal of Personality and Social Psychology}, 107(2), 276--299.

\bibitem[Metaculus(2023)]{Metaculus2023}
Metaculus (2023).
\newblock Metaculus track record.
\newblock \url{https://www.metaculus.com/questions/track-record/}.

\bibitem[Page(2007)]{Page2007}
Page, S.~E. (2007).
\newblock \emph{The Difference: How the Power of Diversity Creates Better Groups, Firms, Schools, and Societies}.
\newblock Princeton University Press.

\bibitem[Schoenegger et~al.(2024)]{Schoenegger2024}
Schoenegger, P., Park, P., Karger, E., \& Tetlock, P.~E. (2024).
\newblock AI-augmented predictions: LLM assistants improve human forecasting accuracy.
\newblock \emph{arXiv preprint arXiv:2402.07862}.

\bibitem[Simmons et~al.(2011)]{Simmons2011}
Simmons, J.~P., Nelson, L.~D., Galak, J., \& Frederick, S. (2011).
\newblock Intuitive biases in choice versus estimation: Implications for the wisdom of crowds.
\newblock \emph{Journal of Consumer Research}, 38(1), 1--15.

\bibitem[Surowiecki(2004)]{Surowiecki2004}
Surowiecki, J. (2004).
\newblock \emph{The Wisdom of Crowds}.
\newblock Doubleday.

\bibitem[Toyokawa et~al.(2019)]{Toyokawa2019}
Toyokawa, W., Whalen, A., \& Laland, K.~N. (2019).
\newblock Social learning strategies regulate the wisdom and madness of interactive crowds.
\newblock \emph{Nature Human Behaviour}, 3(2), 183--193.

\bibitem[Zhu et~al.(2000)]{Zhu2000}
Zhu, Y., Chen, H., Fan, J., et~al. (2000).
\newblock Genetic diversity and disease control in rice.
\newblock \emph{Nature}, 406(6797), 718--722.

\end{thebibliography}
\end{document}